\documentclass{iaus}
\usepackage{graphicx}
\usepackage[breaklinks]{hyperref}

\title[] 
{Single mode, extreme precision Doppler spectrographs}

\author[C. Schwab, S.~G. Leon-Saval, C.~H. Betters, J. Bland-Hawthorn, S. Mahadevan]   
{Christian Schwab$^1$, Sergio~G. Leon-Saval$^2$, Christopher~H. Betters$^2$, Joss Bland-Hawthorn$^2$, Suvrath Mahadevan$^3$}

\affiliation{$^1$Department of Astronomy, Yale University, New Haven, USA
\\ email: {\tt christian.schwab@yale.edu} \\[\affilskip]
$^2$Institute of Photonics \& Optical Science, University of Sydney, Sydney, Australia 
\\email: {\tt sergio.leon-saval@sydney.edu.au}, 
\\email: {\tt c.betters@physics.usyd.edu.au},
\\email: {\tt  jbh@physics.usyd.edu.au}, 
\\$^3$Department of Astronomy, Pennsylvania State University, State College, USA
\\email: {\tt suvrath@astro.psu.edu}}

\pubyear{2012}
\pagerange{1-4}
\begin{document}

\maketitle

\begin{abstract}
The `holy grail' of exoplanet research today is the detection of an earth-like planet: a rocky planet in the habitable zone around a main-sequence star. Extremely precise Doppler spectroscopy is an indispensable tool to find and characterize earth-like planets; however, to find these planets around solar-type stars, we need nearly one order of magnitude better radial velocity (RV) precision than the best current spectrographs provide.  Recent developments in astrophotonics (\cite{Bland-Hawthorn2006}, \cite{Bland-Hawthorn2010}) and adaptive optics (AO) enable single mode fiber (SMF) fed, high resolution spectrographs, which can realize the next step in precision. SMF feeds have intrinsic advantages over multimode fiber or slit coupled spectrographs: The intensity distribution at the fiber exit is extremely stable, and as a result the line spread function of a well-designed spectrograph is fully decoupled from input coupling conditions, like guiding or seeing variations \cite{Ihle2010}. Modal noise, a limiting factor in current multimode fiber fed instruments \cite{Baudrand2001}, can be eliminated by proper design, and the diffraction limited input to the spectrograph allows for very compact instrument designs, which provide excellent optomechanical stability. A SMF is the ideal interface for new, very precise wavelength calibrators, like laser frequency combs (\cite{Steinmetz2008, Osterman2012}), or SMF based Fabry-Perot Etalons \cite{Halverson2012}. At near infrared wavelengths, these technologies are ready to be implemented in on-sky instruments, or already in use. We discuss a novel concept for such a spectrograph. 

\end{abstract}

\firstsection 
\section{Key astrophotonic and AO technologies}

\begin{figure}[t]
\begin{minipage}[b]{0.45\linewidth}
\centering
\includegraphics[width=\textwidth]{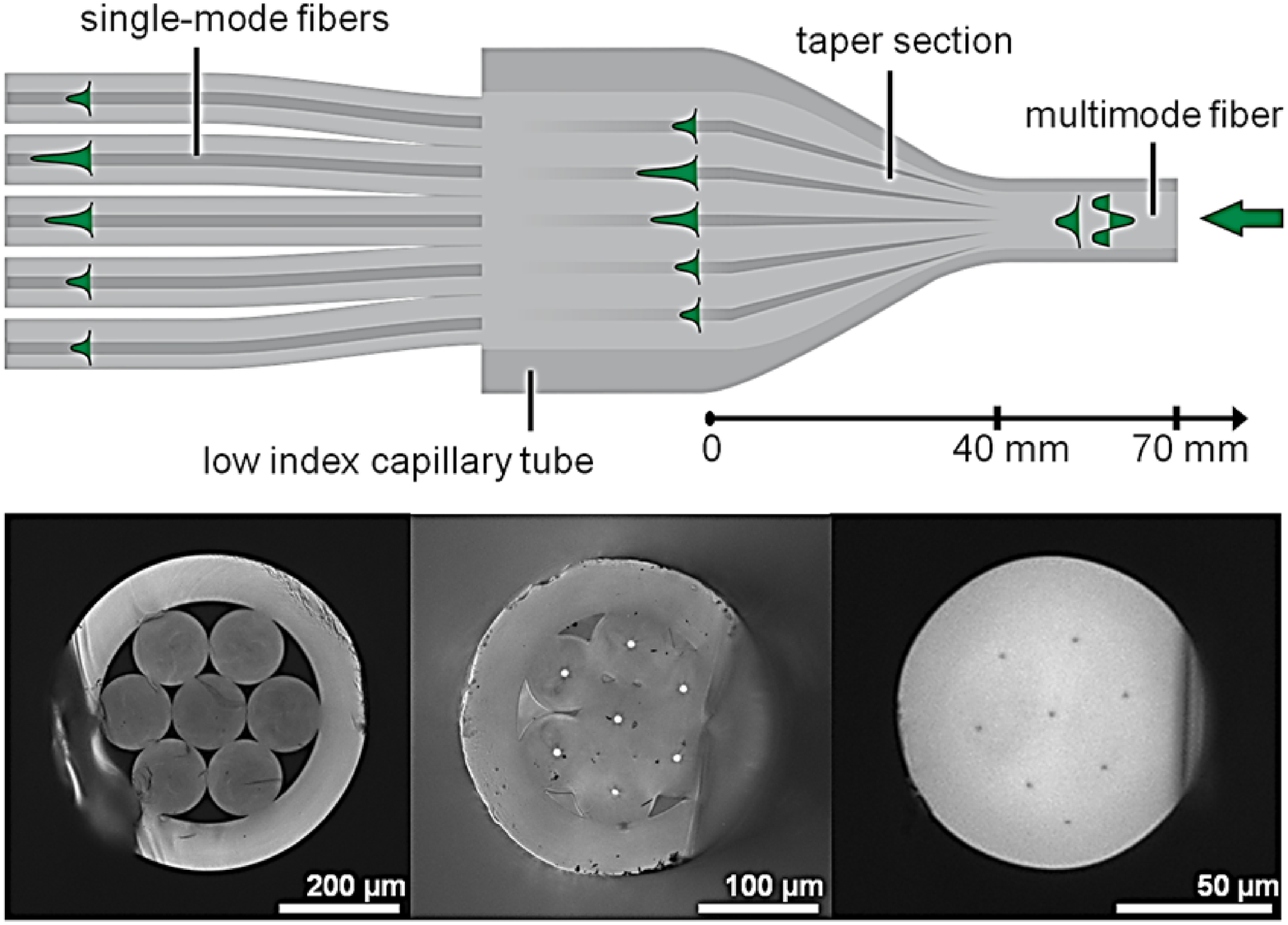}
\caption{(top panel) Schematic illustration of the photonic lantern. (bottom panel) Microscope pictures at different positions along the down taper transition.}
\label{fig:figure1}
\end{minipage}
\hspace{0.5cm}
\begin{minipage}[b]{0.45\linewidth}
\centering
\includegraphics[width=\textwidth]{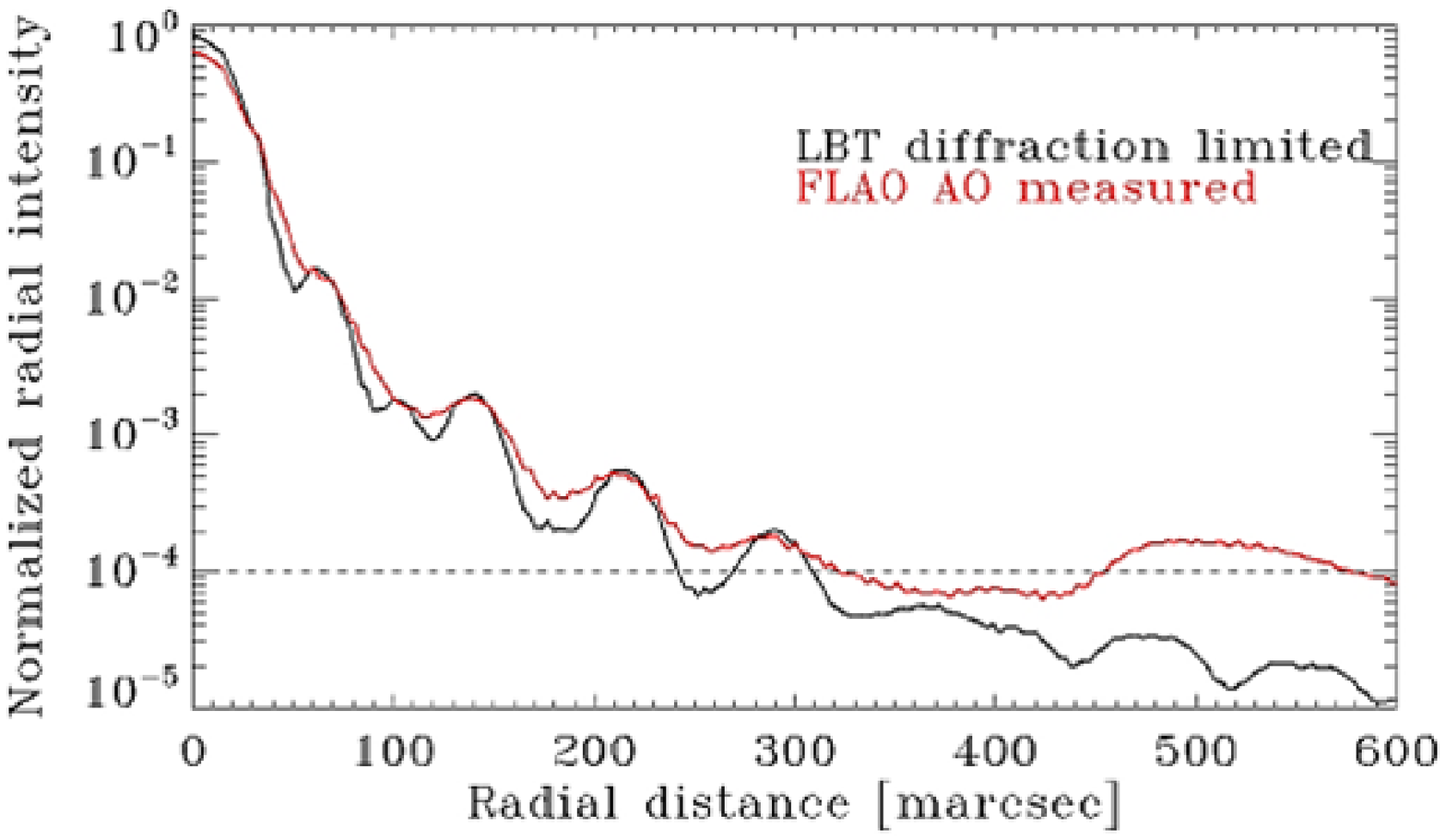}
\caption{High-order AO-corrected PSF in H-band using a 6.5 mag star  under median seeing conditions, obtained with the LBT AO system \cite[(Esposito et al. 2011)]{Esposito2011}.}
\label{fig:figure2}
\end{minipage}
\end{figure}

A so-called photonic lantern is a device which converts a multimode fiber input into a number of single mode fiber outputs \cite{SLS2010}(see Fig. \ref{fig:figure1}). If the number of modes in the multimode fiber is preserved by coupling to the same number of single mode fibers, the conversion can be very efficient;  $>90$\% throughput has been demonstrated by \cite{Noordegraaf2010}. This technology alleviates the problem of direct coupling to a SMF, which easily leads to unacceptable losses even in the case of highly corrected point spread functions \cite{Coude2000}. The spatial size of a seeing limited star necessitates the use of large multimode fibers for efficient coupling, typically around 200 microns diameter at f/5 for an 8m class telescope. This corresponds to 878 vector modes, or 439 fibers, at a wavelength of 1.5 micron, and exceeds the current manufacturing limits for photonic lanterns. Furthermore, as each SMF output needs to be recorded separately, such a spectrograph would require an enormous number of detector pixels. However, the advent of powerful AO systems which achieve excellent correction on bright natural guide stars enables the use of small, few-mode lanterns. \cite{Esposito2011} demonstrated Strehl ratios in excess of 85\% in H-band with the large binocular telescope (LBT) and its adaptive secondary mirror (Fig. \ref{fig:figure2}). A photonic lantern with seven output fibers in H-band has a multimode end corresponding to 90 milliarcseconds diameter at the LBT, and would enclose the AO-corrected point spread function (PSF) to the first minimum in the diffraction pattern. As a consequence, we believe that photonic lanterns can efficiently couple a diffraction limited spectrograph to a large telescope equipped with a powerful AO system. For a high precision radial velocity program, the bright science target stars provide ideal guide stars for a natural guide star AO system.

\section{Photonic lantern fed, high resolution Echelle spectrograph}

\begin{figure}[t]
\begin{center}
 \includegraphics[width=\textwidth]{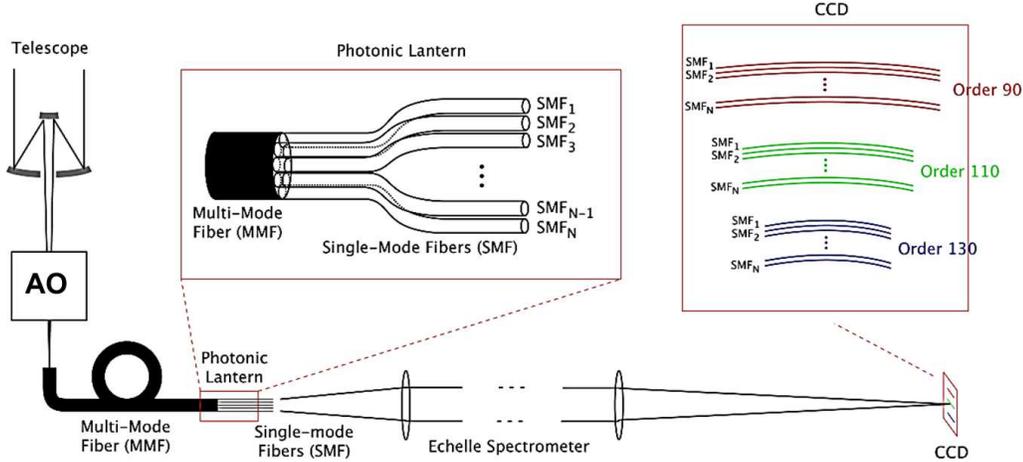} 
 \caption{Schematic layout of our spectrograph concept. A photonic lantern couples to the telescope via an AO system. The single mode fibers at the output provide diffraction limited sources to a compact Echelle spectrograph. The cross dispersion is chosen such that the spectra of the N fibers are clearly separated on the detector. Each Echelle order consists of N spectra.}
   \label{fig:figure3}
\end{center}
\end{figure}

The SMF fiber output from the photonic lantern provides a diffraction limited entrance aperture for the spectrograph (Fig. \ref{fig:figure3}). In this case, the image size (or line width) corresponds to the PSF of the spectrograph optics, and for a well-corrected optical design is determined by diffraction on the spectrograph pupil stop. Very high resolution can be achieved with small beam diameters and grating dimensions. With an R4 (76 degrees blaze angle) grating, $R>100.000$ is possible with a 25 mm diameter beam. Further, to sample the spectrum with common detectors, the focal ratio of the spectrograph camera has to be very slow, to match the point spread function to the pixel size. The combination of slow camera and small optics opens up new opportunities for the optical design that cannot be used for classical spectrographs. It is important to note that the spectra from all fibers need to be recorded separately, so as not to introduce modal noise. As a result, the SMFs can be lined up either in a linear fashion, or in a more compact arrangement, for example in a slightly rotated hexagonal pattern as suggested by \cite{SLS2012}, resulting in a constant offset between the fibers in dispersion direction. The optical design has to provide enough cross dispersion to separate the spectra on the detector.   Another advantage of the compact size is that the spectrograph bench can be mechanically and thermally stabilized to an unprecedented degree, compared with classical spectrographs, and easily evacuated. The small optics can be made to stricter tolerances than large substrates, keeping cost down at the same time. This helps to further suppress any variations in the spectrum due to the spectrograph optics itself. 

\section{Breadboard testing a single mode Echelle spectrograph}

\begin{figure}[htb]
\centering
\includegraphics[width=\textwidth]{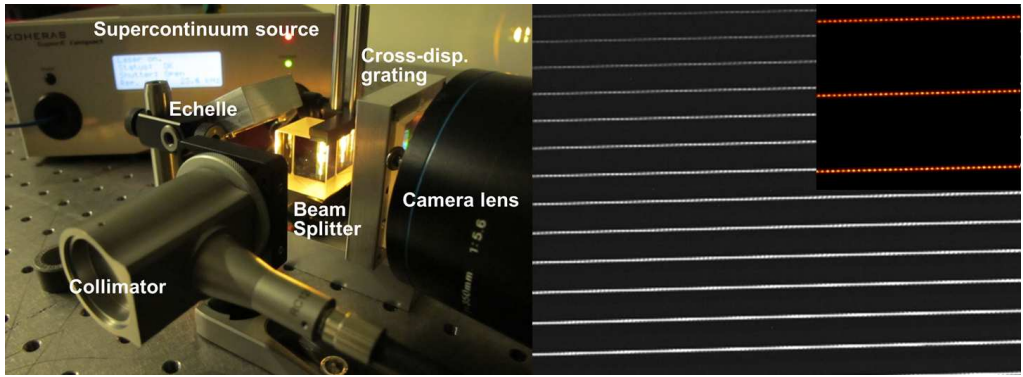}
\caption{Left: Photo of a breadboard spectrograph.  Right: Example of a single mode spectrum recorded with our breadboard setup. We inserted a fused silica solid Etalon with a free spectral range of 0.5 cm$^{-1}$ from Lightmachinery. In the magnified insert, it can be seen that the Etalon peaks are well defined and separated.}
\label{fig:figure4}
\end{figure}

We set up a single mode Echelle spectrograph based on commercial off-the-shelf components to verify line shape and resolution, and to test the integration of a photonic lantern. We chose to test the concept in the visible wavelength region due to the better availability of optics and detectors.  The breadboard shown in Fig. \ref{fig:figure4} is very simple and compact, yet it provides excellent resolution. We used a 25 $\times$ 50 mm, 31.6 g/mm R2 Echelle grating, with a theoretical resolving power close to 60.000. We employ a reflective, one inch fiber collimator based on an off-axis parabola. The light is directed towards the R2 grating with the help of a non-polarizing beamsplitter cube, which allows us to use the grating in true Littrow configuration. The cross disperser is a 300 g/mm transmission grating; the dispersion is high enough to fit a lantern with 7 output cores between the orders. The camera is a commercial, apochromatic telephoto lens with 350 mm focal length, coupled to CCD camera based on a front illuminated Kodak CCD that has 5.4 micron pixels. Fig. \ref{fig:figure4} is an Echelle spectrum of a supercontinuum source, centered at 550 nm, taken with a single SMF. The inset shows the easily resolved peaks of a Fabry-Perot etalon. The peak to peak distance corresponds to a resolution of $\sim$37.000.

\section{Calibration Sources}
The use of a single-mode fiber presents some challenges for a calibration source. Th-Ar \cite{Baranne1996} or U-Ne \cite{Redman2012} lamps that are used in optical and NIR are not suitable sources since their extended cathodes do not couple well to the single mode fibers. Laser frequency combs (e.g., \cite{Ycas2012}, \cite{Steinmetz2008}, \cite{Osterman2012}) offer one possibility for the highest precision applications. Commercially available supercontinuum white light sources enable fiber gas cells \cite{Mahadevan2009} or fiber-Fabry-Perot devices \cite{Halverson2012} to be used effectively with single-mode spectrographs. Some combination of these sources will satisfy most needs, ranging from the need for the highest precision (5-10 cm/s) to having inexpensive commercially available sources (1-5 m/s precision) to enable such AO fed spectrograph systems to be built for a fraction of the cost of large evacuated spectrographs.

\section{Summary}
We are developing a concept for an ultra-stable single mode spectrograph that will be used for high precision Doppler searches for extrasolar planets on large telescopes. To couple the spectrograph to the telescope, we aim to use a photonic lantern behind an AO system. We are currently testing components in the lab, and designing an optimized, efficient, and highly stable spectrograph bench with $R>100.000$. \\

SM acknowledges support from NSF Grants AST1006676 and AST1126413. The Center for Exoplanets and Habitable Worlds is supported by the Pennsylvania State University, the Eberly College of Science, and the Pennsylvania Space Grant Consortium.
Data presented herein was obtained during a Sagan Fellowship supported by the National Aeronautics and Space Administration (NASA) under contract with the Jet Propulsion Laboratory (JPL) through Prime Contract No. NAS7-03001. JPL is managed for NASA by the California Institute of Technology. The Fellowship is administered by the NASA Exoplanet Science Institute (NExScI).



\end{document}